\documentclass[epj,twocolumn]{webofc}
\usepackage[varg]{txfonts}   
\newcommand{\roots}       {\ensuremath{\sqrt{s}}\xspace}
\newcommand{\rootsnn}     {\ensuremath{\sqrt{s_\mathrm{NN}}}\xspace}

\newcommand{\pt}          {\ensuremath{p_\mathrm{T}}\xspace}
\newcommand{\mpt}         {\ensuremath{\langle\pt\rangle}\xspace}

\newcommand{\meanpt}      {\ensuremath{M(\pt)}\xspace}
\newcommand{\meanptm}     {\ensuremath{M(\pt)_{m}}\xspace}

\newcommand{\rspt}        {\ensuremath{\sqrt{C_{m}}/\meanptm}\xspace}
\newcommand{\rsptinc}     {\ensuremath{\sqrt{C}/\meanpt}\xspace}

\newcommand{\nacc}        {\ensuremath{N_\mathrm{acc}}\xspace}

\newcommand{\nch}         {\ensuremath{N_\mathrm{ch}}\xspace}

\newcommand{\dndeta}      {\ensuremath{\mathrm{d}\nch/\mathrm{d}\eta}\xspace}
\newcommand{\mdndeta}     {\ensuremath{\langle\dndeta\rangle}\xspace}
\newcommand{\npart}       {\ensuremath{N_{\mathrm{part}}}\xspace}
\newcommand{\mnpart}      {\ensuremath{\langle\npart\rangle}\xspace}
\woctitle{XLIV International Symposium on Multiparticle Dynamics}
\begin{document}
\title{Event-by-event mean $p_{\rm T}$ fluctuations in pp and Pb--Pb collisions at the LHC}
\author{Stefan Heckel\inst{1}\fnsep\thanks{\email{sheckel@ikf.uni-frankfurt.de}} on behalf of the ALICE Collaboration
}
\institute{Institut f\"{u}r Kernphysik, Goethe-Universit\"{a}t, Max-von-Laue-Str. 1, 60438 Frankfurt am Main, Germany
          }
\abstract{%
The ALICE detector at the LHC is used to study the properties of the Quark-Gluon Plasma 
produced in heavy-ion collisions. As a reference measurement, also the analysis of 
proton-proton (pp) collisions is very important.
In the study presented here, event-by-event fluctuations of the mean transverse momentum 
are analysed in pp collisions at \roots~$=$~0.9, 2.76 and 7~TeV, and Pb--Pb collisions
at \rootsnn~$=$~2.76~TeV as a function of the charged-particle multiplicity.
In both systems, dynamical fluctuations beyond the statistical expectation are observed. 
In pp collisions, no significant dependence on collision energy is found, even in 
comparison to inclusive results at much lower collision energies. 
Likewise, central A--A collisions show only little dependence on collision energy. 
The multiplicity dependence observed in peripheral Pb--Pb data is in agreement with that in 
pp collisions. Going to more central Pb--Pb collisions, a clear deviation from this trend is 
found, reaching a significant reduction of the fluctuations in most central collisions.
Comparisons to Monte Carlo event generators show good agreement in pp, but rather large 
differences in Pb--Pb collisions.
}
\maketitle
\section{Introduction}\label{sec:intro}

At LHC energies, not only heavy-ion (A--A) collisions, but also pp collisions produce a large number of particles in the final state. The ALICE detector~\cite{alice} is well suited for the investigation of both systems and can cope in particular with the very large number of particles produced in central Pb--Pb collisions. Its main purpose is the study of the Quark-Gluon Plasma, where the quarks and gluons are believed to exist in a quasi-free state and are recombined into colorless objects at the phase transition to a Hadron Gas.
This phase transition may go along with critical fluctuations of thermodynamic quantities which could be observed via event-by-event fluctuation measurements. These analyses may also reveal collective effects like the onset of thermalization of a system and are therefore proposed for the investigation of the hot and dense matter generated in heavy-ion collisions~\cite{jeonkoch,shuryak,stephanov1,stephanov2}.

Recently, ALICE has measured event-by-event fluctuations of the mean transverse momentum $\left(  \left<p_{\rm T}\right> \right)$ of final-state charged particles as a function of the average charged-particle multiplicity density $\left( \mdndeta \right)$~\cite{aliceptfluc}. Many kinds of correlations among the transverse momenta of the final-state particles may lead to such fluctuations, e.g.\,jets, resonance decays and quantum correlations. These are already present in pp collisions, which can hence serve as a model-independent baseline of the investigation of additional effects in A--A collisions. We present results for pp collisions at \roots~$=$~0.9, 2.76 and 7~TeV, and Pb--Pb collisions at \rootsnn~$=$~2.76~TeV together with a comparison of both systems. Furthermore, we compare to previous measurements at lower collision energies and to several Monte Carlo (MC) event generators.

\section{ALICE detector and analysis details}\label{sec:analysis}

The results presented are based on data measured at the CERN Large Hadron Collider (LHC)~\cite{lhc} with the ALICE detector~\cite{alice} taking into account $19\times10^6$ Pb--Pb events at \rootsnn~$=$~2.76~TeV, and $6.9\times10^6$, $66\times10^6$ and $290\times10^6$ pp events at \roots~$=$~0.9, 2.76 and 7 TeV, respectively.

Charged-particle tracking is performed with the Time Projection Chamber (TPC)~\cite{tpc} in the pseudo-rapidity range $|\eta|<0.8$. The transverse momentum range is restricted to $0.15<p_{\rm T}<2$~GeV/$c$, both to guarantee a good tracking efficiency and momentum resolution and because the focus of this study is on bulk particle production. With the upper \pt cut, high \pt particles originating from initial hard scatterings are reduced.

For the determination of the primary vertex, the Inner Tracking System (ITS) is used in addition to the TPC. Events are only accepted in the analysis, when at least one charged-particle track is contributing to the primary vertex reconstruction, which itself has to be within $\pm10$~cm from the nominal interaction point along the beam direction to ensure a uniform pseudo-rapidity acceptance within the TPC. In Pb--Pb collisions, an additional requirement of at least 10 reconstructed tracks inside the acceptance removes the largest fraction of non-hadronic events. The two forward scintillator systems VZERO-A ($2.8 < \eta < 5.1$) and VZERO-C ($-3.7 < \eta < -1.7$) are used for the centrality determination in Pb--Pb as outlined in~\cite{dndeta}.

On the level of single charged particles, a set of different track selection criteria assures a good quality of the accepted tracks. The number of accepted tracks in one event within the $\eta$ and \pt ranges given above is denoted as \nacc.

The determination of a quantity like the mean transverse momentum \mpt on an event-by-event level is subject to efficiency loss of particle tracks as well as to the finite momentum resolution. Therefore, the event-by-event mean transverse momentum
is approximated by the mean value $M_{\rm EbE}(p_{\rm T})_k$ of the transverse momenta $p_{{\rm T},i}$ of the $N_{\rm acc},_k$ accepted charged particles in event $k$:
\begin{equation}
  M_{\rm EbE}(p_{\rm T})_k=\frac{1}{N_{\rm acc},_k} \sum_{i=1}^{N_{\rm acc},_k} p_{{\rm T},i} \ .
\end{equation}
In general, both statistical and dynamical fluctuations of $M_{\rm EbE}(p_{\rm T})_k$ are present. A measure for only the dynamical contribution is the two-particle correlator $C = \langle \Delta p_{{\rm T},i}, \Delta p_{{\rm T},j} \rangle$ \cite{ceres-2,star-2,voloshin}. In this analysis, the correlator $C_{m}$ is used, which is defined in multiplicity classes $m$. The mean transverse momentum of all tracks in all events of class $m$ is denoted as \meanptm and defined as
\begin{align}
  M(p_{\rm T})_m &= \frac{1}{\sum_{k=1}^{n_{{\rm ev},m}}N_{\rm acc},_k} \sum_{k=1}^{n_{{\rm ev},m}}\sum_{i=1}^{N_{\rm acc},_k} p_{{\rm T},i} \nonumber \\ &=\frac{1}{\sum_{k=1}^{n_{{\rm ev},m}}N_{\rm acc},_k} \sum_{k=1}^{n_{{\rm ev},m}} N_{\rm acc},_k \cdot M_{\rm EbE}(p_{\rm T})_k \ .
\end{align}
Furthermore, $n_{{\rm ev},m}$ is the number of events in multiplicity class $m$ and $N_{k}^{\rm pairs} = 0.5 \cdot N_{\rm acc},_{k} \cdot (N_{\rm acc},_{k}-1)$ is the number of particle pairs in event $k$.
With this, the two-particle correlator $C_{m}$, which is the mean of covariances of all pairs of particles $i$ and $j$ in the same event 
with respect to the inclusive $M(p_{\rm T})_m$, can be defined as
\begin{align}
  C_m = &\frac{1}{\sum_{k=1}^{n_{{\rm ev},m}}{N_{k}^{\rm pairs}}} \cdot \sum_{k=1}^{n_{{\rm ev},m}} \sum_{i=1}^{N_{\rm acc},_{k}} \sum_{j=i+1}^{N_{\rm acc},_{k}} \nonumber \\ &(p_{{\rm T},i} - M(p_{\rm T})_m ) \cdot (p_{{\rm T},j} - M(p_{\rm T})_m) \ .
  \label{eq:correlator}
\end{align}
By construction, $C_m$ vanishes, if only statistical fluctuations are present. The dimensionless ratio \rspt is used for the presentation of the results which quantifies the strength of the dynamical fluctuations in units of the average transverse momentum $M(p_{\rm T})_m$.

\section{Results in pp collisions}\label{sec:respp}

\begin{figure}
\centering
	\includegraphics[width=0.48\textwidth]{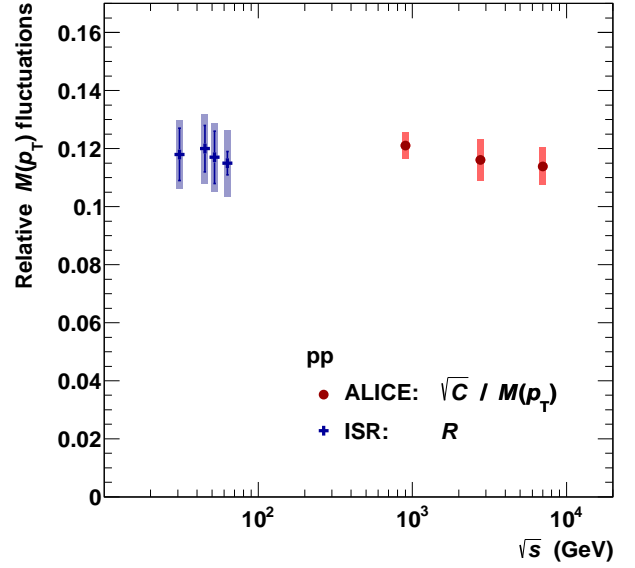}
	\caption{Relative dynamical mean transverse momentum fluctuations
	in pp collisions as a function of \roots. The ALICE results for \rsptinc~are compared
	to the quantity $R$ measured at the ISR (see text and~\cite{isr}).}
	\label{fig:roots}
\end{figure}

The inclusive relative dynamical fluctuation \rsptinc, which is not subdivided into multiplicity classes $m$, is shown as a function of \roots in figure~\ref{fig:roots} for pp collisions at \roots~$=$~0.9, 2.76 and 7~TeV. Significant dynamical \meanpt fluctuations are observed with a magnitude of about 12~\% of \meanpt, independent of the collision energy.

The ALICE results are compared to a similar quantity $R$ measured by the Split Field Magnet (SFM) detector at the ISR in pp collisions at much lower collision energies of \roots~$=$~30.8, 45, 52, and 63 GeV~\cite{isr}. $R$ is determined by extrapolating the multiplicity-dependent dispersion $D(  M_{\rm EbE}(p_{\rm T})_k)$ to infinite multiplicity and normalization by the inclusive mean transverse momentum. This is an alternative estimate of dynamical mean transverse momentum fluctuations. The comparison of ALICE and ISR results at collision energies separated by two orders of magnitude does not reveal any significant dependence of the fluctuations on the collision energy.

\begin{figure}
\centering
	\includegraphics[width=0.48\textwidth]{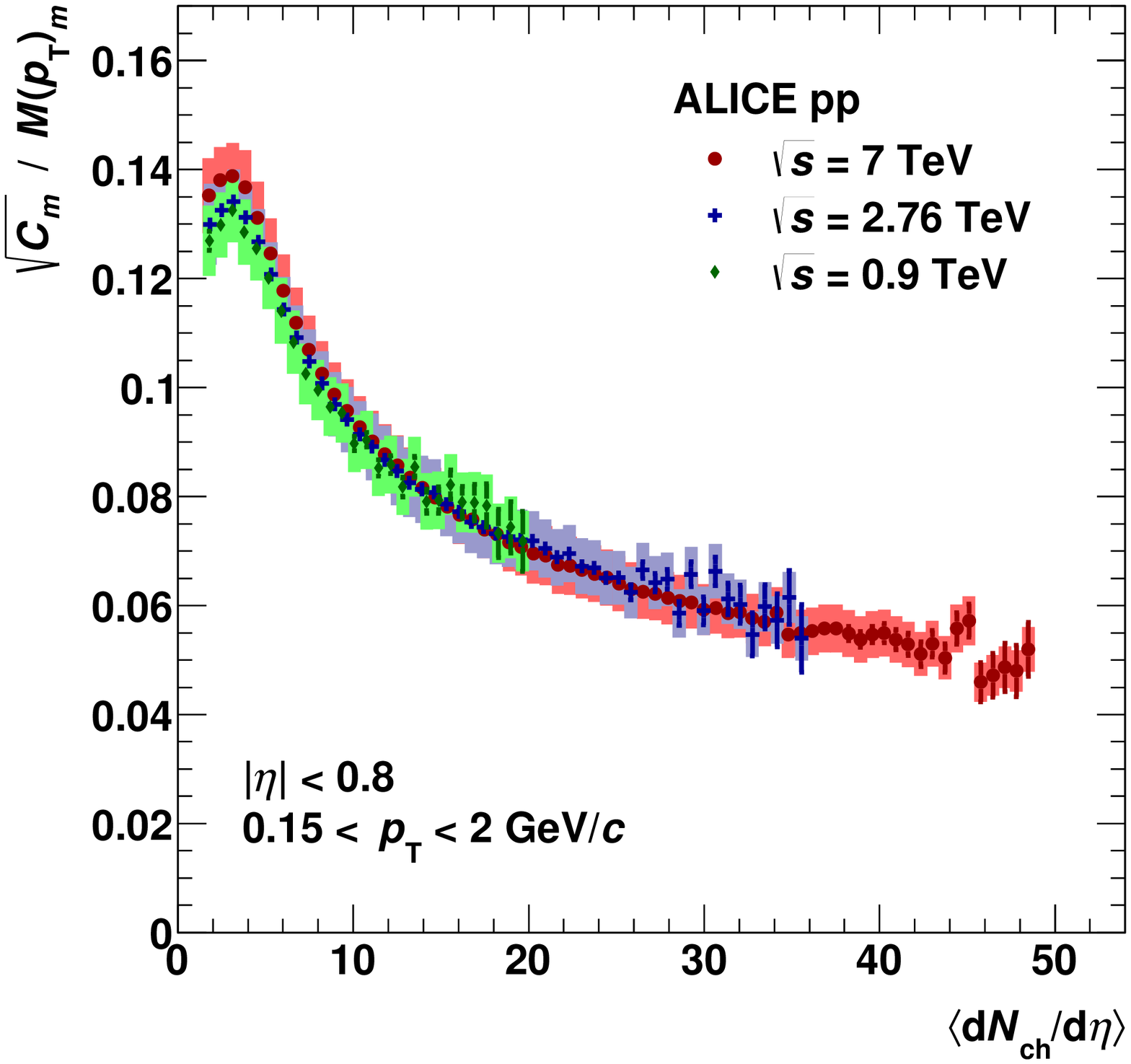}
	\caption{Relative fluctuation \rspt~as a function of \mdndeta~in pp collisions 
	at \roots~$=$~0.9, 2.76 and 7~TeV.}
	\label{fig:pp-corr}
\end{figure}
\begin{figure*}[t]
\centering
	\includegraphics[width=0.48\textwidth]{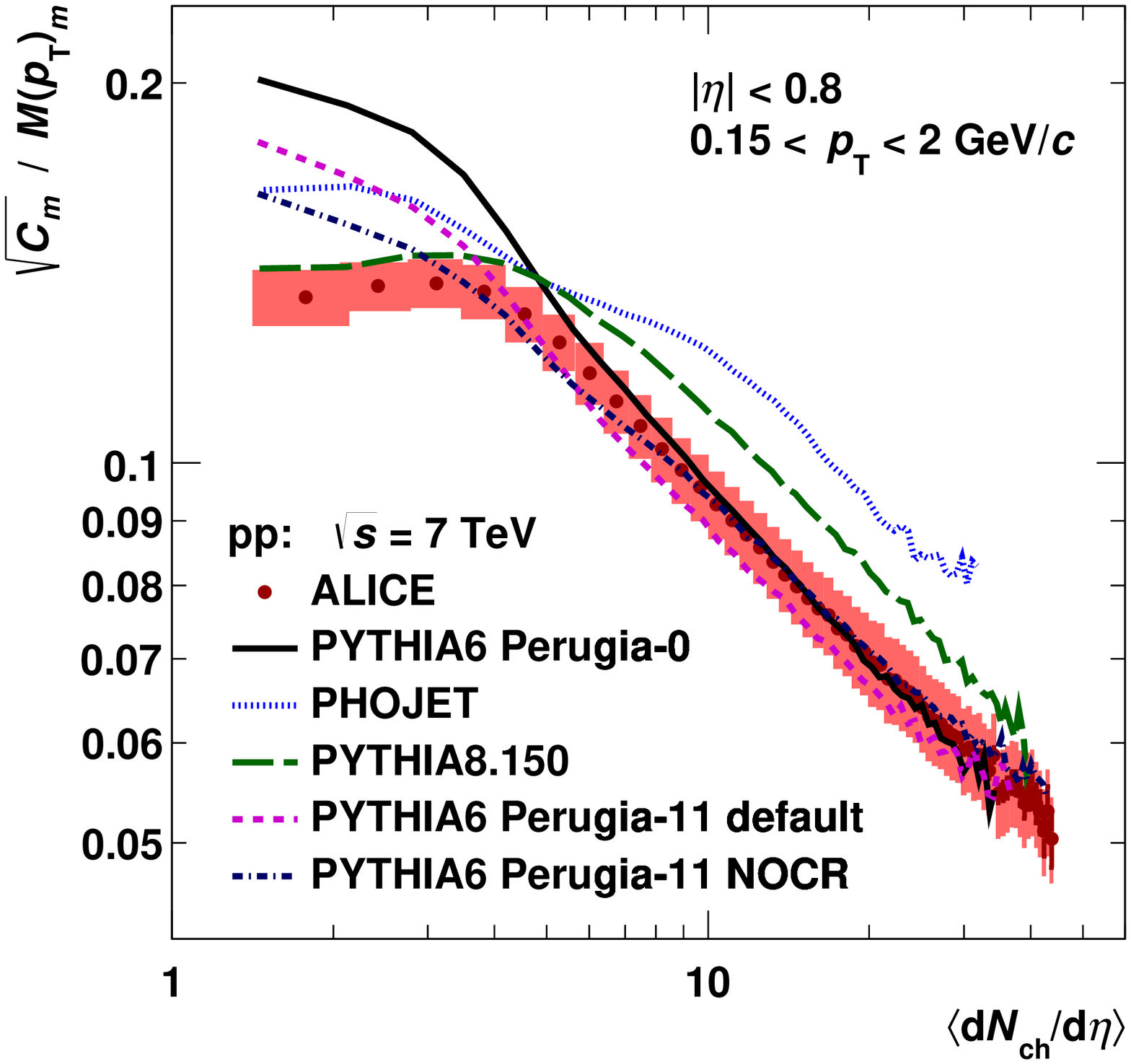}
	\includegraphics[width=0.48\textwidth]{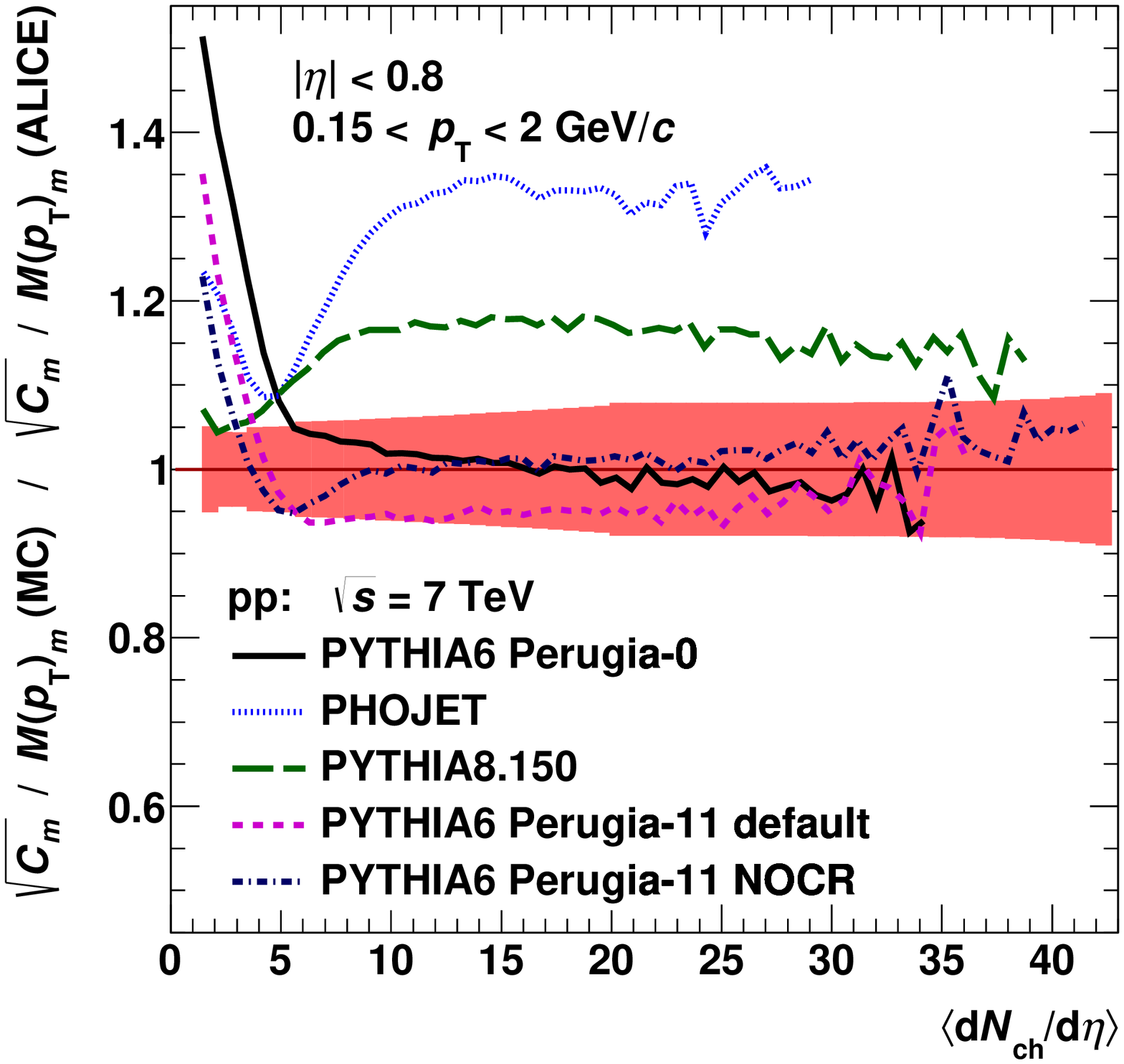}
	\caption{Left: Relative dynamical fluctuation \rspt~for data and different 
	event generators in pp collisions at \roots~$=$~7~TeV as a function of \mdndeta. 
	Right: Ratio models to data. The red error band indicates the statistical and systematic 
	data uncertainties added in quadrature.}
	\label{fig:mc-pp}
\end{figure*}

At LHC energies and with the large statistics collected by ALICE so far, it is possible to study event-by-event observables in a more differential way. As a first step, the ALICE data are analysed as a function of the average charged-particle multiplicity density \mdndeta in pp collisions at \roots~$=$~0.9, 2.76 and 7~TeV. In figure~\ref{fig:pp-corr}, the relative dynamical fluctuation \rspt is presented as a function of \mdndeta for all three collision energies. Dynamical fluctuations are observed over the entire multiplicity range. At low multiplicities, the fluctuation strength reaches values of 12--14~\%, similar to the inclusive results, but with increasing multiplicity the fluctuations decrease to about 5~\% at the highest multiplicities. It is the first time, that \meanpt fluctuations in pp collisions are analysed as a function of multiplicity and that this substructure underlying the inclusive values is observed. The independence of the results on the collision energy is preserved when going to the multiplicity dependent results.

In pp collisions at very high energies as reached at the LHC the occurence of multiple hard scatterings in a single collision may be possible, due to the large number of partons within the ultra-relativistically moving particles. High-multiplicity events in pp collisions at LHC energies are believed to be driven by such multi-parton interactions (MPIs)~\cite{mpi}. These multiple hard scatterings are independent processes in the initial state of the collision. However, the particles originating from different hard scatterings may be recombined via the color-reconnection mechanism, which could lead to correlations in the final state.

These mechanisms cannot be studied directly in the experimental results. Instead, a comparison with several Monte Carlo (MC) event generators is performed in pp collisions at \roots~$=$~7~TeV. PHOJET~\cite{phojet} and several versions and tunes of PYTHIA~\cite{skands,pythia} are used, in particular PYTHIA6 with the Perugia-0 and Perugia-11 tunes and PYTHIA8. In the case of PYTHIA6 Perugia-11, the default tune is compared with another version, where the color-reconnection mechanism is switched off. In PYTHIA, color reconnections are responsible for the increase of \meanpt with multiplicity~\cite{cr,mptvsnch} which hence could lead to interesting results in the case of \meanpt fluctuations.

The relative dynamical fluctuation \rspt as a function of the charged-particle multiplicity density \mdndeta from ALICE in pp collisions at \roots~$=$~7~TeV is compared to the MC generators in figure~\ref{fig:mc-pp} (left panel) and the ratio of the MC generators to ALICE data is shown in figure~\ref{fig:mc-pp} (right panel). \rspt as function of \mdndeta shows a power-law behavior in data which is reproduced by most of the MC generators. A power-law fit to the ALICE data in the interval $5<$\,\mdndeta\,$<30$ yields $b= -0.431 \pm 0.001$\,(stat.)\,$\pm 0.021$\,(syst.). A simple superposition scenario, as would be expected in the case of independent MPIs, would lead to a power-law index of $b=-0.5$, which is clearly not found. An interesting observation is, that both PYTHIA6 Perugia-0 tunes -- with and without the color-reconnection mechanism -- show a very similar behavior and are both in agreement with the data. Unlike the behavior in \meanpt, where color reconnections are necessary to describe the increase with multiplicity, in \meanpt fluctuations the color-reconnection mechanism seems to have no significant influence.

\section{From pp to heavy-ion collisions}\label{sec:pptohi}

The analysis of event-by-event \meanpt fluctuations in pp collisions already shows some interesting behavior and effects.  Relative dynamical fluctuations decreasing with increasing multiplicity are found. The comparison to MC generators shows no hint for a significant role of the mechanism of color reconnections. The influence of multi-parton interactions is subject to future studies.

Going on to the investigation of heavy-ion collisions it is interesting to study, 
whether the trend in pp collisions may serve as a baseline for heavy ions, and to search for effects beyond those observed in pp. In heavy-ion collisions, 
a lot of nucleons collide, leading to a large number of initial hard scatterings.  
It may also be possible, that in some of those collisions multi-parton interactions take place, but this effect is expected to be negligible with respect to the large number of nucleon--nucleon collisions. Other effects unique to heavy ions may originate from the phase transition from a Quark-Gluon Plasma to a Hadron Gas, from fluctuations in the initial state of the collisions and from collective effects like the thermalization of the system.

\begin{figure}
\centering
	\includegraphics[width=0.48\textwidth]{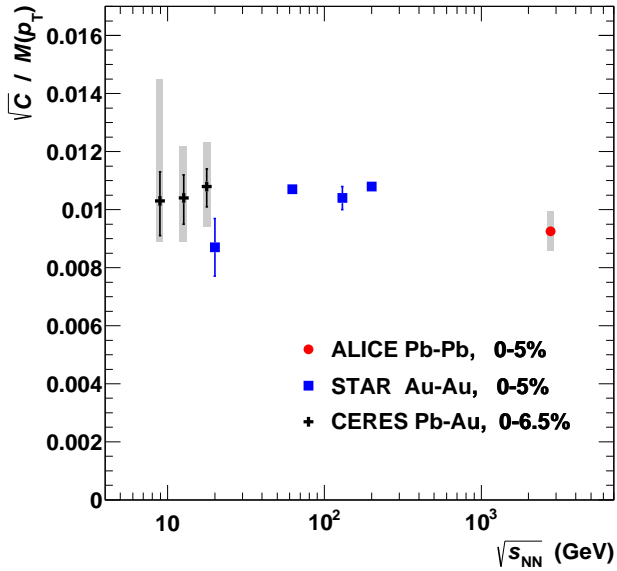}
	\caption{Mean transverse momentum fluctuations in central heavy-ion collisions 
	as a function of \rootsnn. 
	The ALICE data point is compared to data from the CERES~\cite{ceres-2} 
	and STAR~\cite{star-2} experiments. 
	For STAR only statistical uncertainties are available.}
	\label{fig:energy-AA}
\end{figure}

As a first step, most central Pb--Pb collisions at \rootsnn~$=$~2.76~TeV are analysed and compared to results in A--A collisions at lower collision energies measured by the CERES~\cite{ceres-2} and STAR~\cite{star-2} experiments, which is shown in figure~\ref{fig:energy-AA}. As in pp, significant dynamical fluctuations and no significant dependence on the collision energy are observed. The scale of the fluctuations is about 1\% of \meanpt, which is about one order of magnitude smaller than in inclusive pp collisions.

\begin{figure}
\centering
	\includegraphics[width=0.48\textwidth]{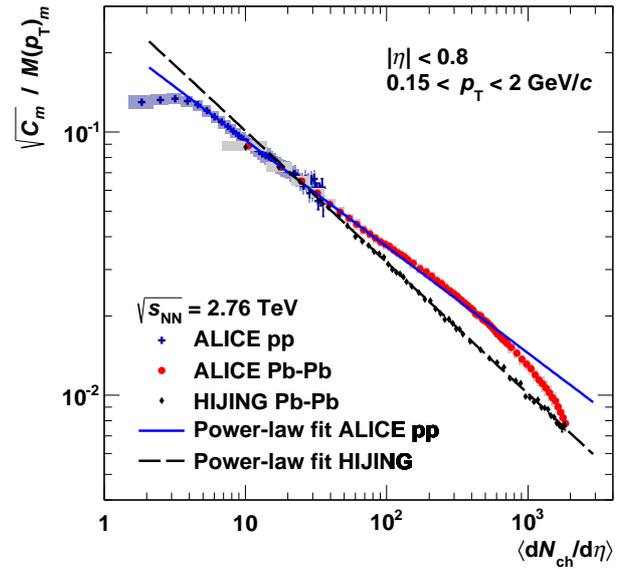}
	\caption{Relative dynamical fluctuation \rspt~as a function of \mdndeta~in pp and Pb--Pb collisions 
	at \rootsnn~$=$~2.76~TeV. Also shown are results from HIJING and power-law fits to pp (solid line) 
	and HIJING (dashed line) (see text).}
	\label{fig:pp-PbPb}
\end{figure}

Figure~\ref{fig:pp-PbPb} presents the relative dynamical fluctuation \rspt as a function of the average charged-particle multiplicity density \mdndeta in pp and Pb--Pb collisions at \rootsnn~$=$~2.76~TeV measured by ALICE. A strong decrease of the fluctuation strength with increasing multiplicity is, as in pp, as well observed in Pb--Pb collisions. Furthermore, the slope of this decrease shows a power-law behavior in peripheral Pb--Pb collisions, which is in very good agreement with the extrapolation of a power-law fit to pp data at \roots~$=$~2.76~TeV in the interval $5<$\,\mdndeta\,$<25$, with $b= -0.405 \pm 0.002$\,(stat.)\,$\pm 0.036$\,(syst.). This agreement holds up to \mdndeta~$\approx$~100, from where on a slight enhancement over the pp extrapolation is observed up to \mdndeta~$\approx$~500. Going to even higher multiplicities, corresponding to collision centralities < 40\%, a strong decrease of the fluctuations below the pp extrapolation is found.

In addition, the result of a HIJING~\cite{hijing} MC simulation is shown in figure~\ref{fig:pp-PbPb}. The HIJING version is 1.36 and no jet quenching is taken into account. Like for the pp data, also for the HIJING simulation a power-law fit is performed, taking the interval $30<$\,\mdndeta\,$<1500$. An exponent $b= -0.499 \pm 0.003$\,(stat.)\,$\pm 0.005$\,(syst.) is found which is in agreement with a simple superposition expectation of $b=-0.5$. The HIJING results are described very well by this fit with the exception of very low multiplicities. Obviously, HIJING cannot describe the behavior observed in Pb--Pb data.

\section{Results in Pb--Pb collisions}\label{sec:respbpb}

One of the effects, which may be important in heavy-ion collisions and is not present in pp collisions, is the fluctuation of the initial conditions. This may be related to event-by-event fluctuations of radial flow and azimuthal asymmetries, which could have an influence on the event-by-event \meanpt fluctuations, although the corresponding event-averaged quantities are not expected to give rise to strong \meanpt fluctuations in azimuthally symmetric detectors~\cite{ceres-2,phenix-2}.

\begin{figure}
\centering
	\includegraphics[width=0.48\textwidth]{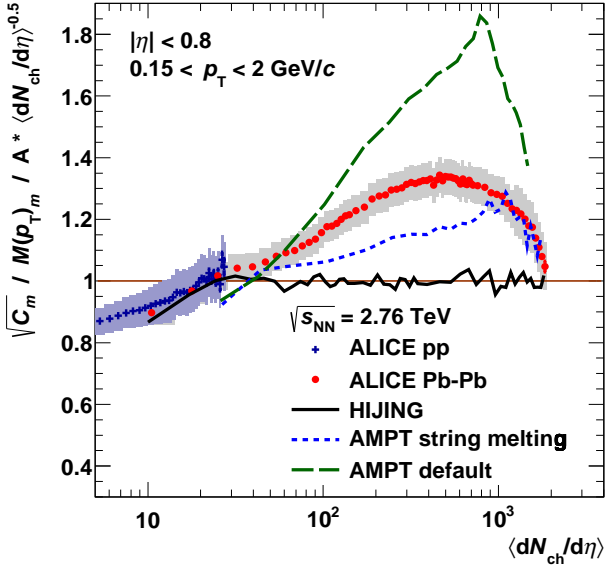}
	\caption{Relative dynamical fluctuation \rspt normalized to \mdndeta$^{-0.5}$ (see text) 
	as a function of \mdndeta in pp and Pb--Pb collisions at \rootsnn~$=$~2.76~TeV. 
	The ALICE data are compared to results from HIJING and AMPT.}
	\label{fig:PbPb-models}
\end{figure}

In figure~\ref{fig:PbPb-models}, a comparison of the pp and Pb--Pb data at \rootsnn~$=$~2.76~TeV to HIJING and two versions of the AMPT~\cite{ampt} simulation, which include collective effects, is presented. The relative dynamical fluctuation \rspt is normalized to the result of a fit of $A\cdot$\mdndeta$^{-0.5}$ to the HIJING simulation in the interval $30<$\,\mdndeta\,$<1500$. HIJING agrees with the data only at very low multiplicities up to \mdndeta~$\approx$~30 and shows a very good agreement with the simple superposition expectation above. The rise of the Pb--Pb data with respect to this expectation as well as the strong decrease towards central events is clearly visible in this representation. Both AMPT calculations agree qualitatively with the behavior observed in the data, but quantitatively all MC generators fail to describe it. The default AMPT calculation shows much larger fluctuations than the data, while the results of the AMPT simulation including the string melting mechanism do not reach the values of the data. In the string melting scenario, partons are not only generated from (hard) mini-jets as in the default version, but also from (soft) strings. In addition, the partons are recombined by a hadronic coalescence scheme after rescattering.

\begin{figure*}
\centering
	\includegraphics[width=0.9\textwidth]{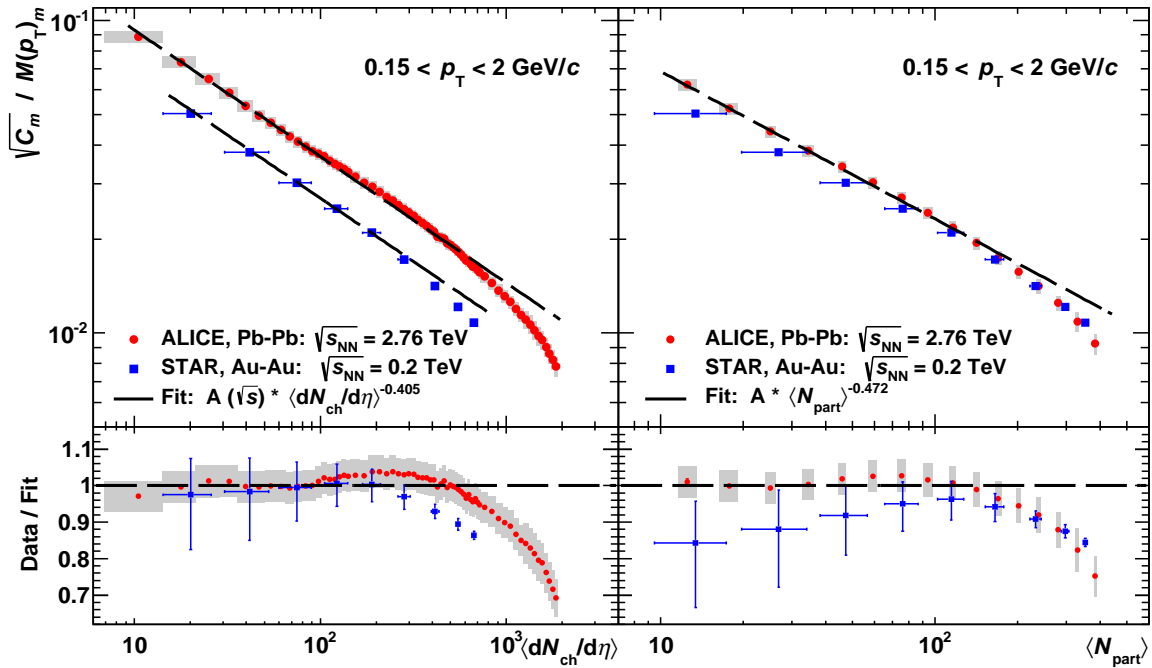}
	\caption{Left: Relative dynamical fluctuation \rspt as a function of \mdndeta in Pb--Pb collisions 
	at \rootsnn~$=$~2.76~TeV from ALICE compared to results from STAR in Au--Au collisions 
	at \rootsnn~$=$~200~GeV~\cite{star-2}. 
	Also shown as dashed lines are results from power-law fits to the data (see text). 
	Right: same data as a function of \mnpart.}
	\label{fig:ALICE-STAR}
\end{figure*}

The ALICE results for \rspt in Pb--Pb collisions at \rootsnn~$=$~2.76~TeV are compared to results by the STAR experiment in Au--Au collisions at \rootsnn~$=$~200~GeV~\cite{star-2} in figure~\ref{fig:ALICE-STAR} as a function of the average charged-particle multiplicity density \mdndeta (left panel) and as a function of the average number of participating nucleons \mnpart (right panel). As a function of \mdndeta, a clear difference of the results is observed. At same multiplicities, a higher fluctuation signal is found in the ALICE data, but also the multiplicity reach is much higher than in STAR. Despite this difference, the STAR data exhibit a very similar scaling with \mdndeta as the ALICE data at much higher energy. In the left panel, also the fit to the ALICE pp results is shown, and a corresponding fit to the STAR data with the same exponent $b=-0.405$. Besides this agreement in the peripheral region, also the deviation of the behavior with an additional decrease of the fluctuations towards central collisions is observed in the STAR measurements, although it is not as pronounced as in the ALICE heavy-ion data. In the right panel of figure~\ref{fig:ALICE-STAR}, i.e. the representation as a function of \mnpart, both measurements are compatible within the rather large experimental uncertainties on \mnpart~in STAR. Here, a power-law fit \rspt\,$\propto$\,\mnpart$^{b}$ is only performed on the ALICE data (in the interval $10<$\,\mnpart\,$<40$), resulting in an exponent $b= -0.472 \pm 0.007$\,(stat.)\,$\pm 0.037$\,(syst.). Both ALICE and STAR data agree with this fit in the peripheral region and show a deviation from the fit at the same \mnpart, i.e.\,the same centralities, which hints to a relation between the measured fluctuation signals and the collision geometry.

\section{Conclusions}\label{sec:conclusions}

Event-by-event mean transverse momentum fluctuations measured in pp collisions at \roots~$=$~0.9, 2.76 and 7~TeV and Pb--Pb collisions at \rootsnn~$=$~2.76~TeV by ALICE at the LHC are presented. In both collisions systems, significant dynamical fluctuations are observed, which decrease with increasing multiplicity. No significant collision energy dependence is found in pp collisions as a function of the average charged-particle multiplicity density \mdndeta and for inclusive results, also in comparison to results at much lower collision energies measured at the ISR. The results in central A--A collisions, compared to lower collision energy data from CERES at SPS and STAR at RHIC, seem to be independent of the collision energy as well.

The trend of relative dynamical fluctuations as a function of \mdndeta observed in pp collision follows a power-law like behavior with an exponent $b$ being different from a simple superposition expectation. An extrapolation of this trend agrees very well with peripheral Pb--Pb data. At higher multiplicities, a slight increase above this pp baseline is followed by a clear additional decrease of the fluctuation strength towards most central collisions. This behavior is similar to the one observed in Au--Au collisions at STAR at much lower collision energies, especially as a function of a geometric quantity like \mnpart, which points to a relation between the fluctuations and the collision geometry.

In pp collisions, the behavior of the measured data is rather well described by a set of different Monte Carlo event generators. Especially, there is no significant difference observed between tunes in PYTHIA6 (Perugia-11) with and without the color reconnection mechanism. Pb--Pb data are quantitatively neither described by HIJING, nor by AMPT. Qualitatively, both AMPT versions under study agree better with the data than HIJING, which points to the importance of including collective effects in the models.
%
\bibliography{PtFlucProceedingsISMD14}
\end{document}